\documentstyle[prd,aps,preprint,tighten]{revtex}

\begin{document}

\title{Dissipative Dynamics of Inflation}

\author{Arjun Berera}

\address{
Department of Physics and Astronomy, University of 
Edinburgh,
Edinburgh, EH9 3JZ, Scotland
\\E-mail:ab@ph.ed.ac.uk}

\maketitle

\abstract{
Dissipative scalar quantum field theory is examined at zero
temperature.  Estimates of radiation production are given.
Relevance of the results to supercooled and warm inflation
are discussed.}

\medskip

To appear in Proceeding, PASCOS-01, 2001

\bigskip

\bigskip

The basic picture of inflationary dynamics centers around a scalar
field often called the inflaton.  During the inflationary
period, the potential energy of this field is pictured to dominate
the energy density of the universe, thereby driving inflation-like
accelerated expansion of the scale factor.  The inflaton field
also is required to interact with other fields, so as to allow
transfer of energy from potential energy into radiation.
Eventually the radiation energy density must dominate
so that inflation can terminate into a standard hot big-bang
radiation dominated regime.  Although ultimately for inflationary
dynamics to fit into a realistic particle physics scheme,
the final models may be more elaborate, it is believed that these
simple inflaton models contain all the essential features that must
be found in any more realistic model.

The most nontrivial aspect of the inflaton models is understanding the
energy transfer dynamics from potential energy to radiation.
A commonly followed picture is that dissipative effects
of the inflaton field can be ignored throughout
the inflation period, thus leading to a supercooled inflationary
regime.  However, from a thermodynamic perspective, this
picture appears very restrictive. The point being, even if
the inflaton were to allow a minuscule fraction
of the energy to be released, say one part in $10^{20}$,
it still would constitute a significant radiation
energy density component in the universe.
{}For example, for inflation with vacuum (i.e. potential) energy
at the GUT scale $\sim 10^{15-16} {\rm GeV}$,
leaking one part in $10^{20}$ of this energy into radiation 
corresponds to a temperature of $10^{11} {\rm GeV}$,
which is nonnegligible.  In fact, the most relevant
lower bound that cosmology places on the temperature after inflation
comes from the success of hot Big-Bang nucleosynthesis,
which thus requires the universe to be within the
radiation dominated regime by $T \stackrel{>}{\sim} 1 {\rm GeV}$.
This limit can be met in the above example by dissipating 
as little as one part in $10^{60}$ of the vacuum energy
into radiation.
Thus, from the perspective of both interacting field theory and
basic notions of equipartition,  it appears to be a highly
tuned requirement of supercooled inflation to prohibit
the inflaton from even such tiny amounts of dissipation.

These considerations have led to examining the possibility
of warm inflation, an inflationary regime in which radiation
also is present.
Warm inflation is comprised of non-isentropic expansion in the
background cosmology \cite{wi} and thermal seeds of density
perturbations \cite{bf2,abnpb,wipert} (see also \cite{moss}).  
During warm inflation,
interactions between the inflaton
and other fields cause
the radiation energy
density to remain substantial due to its constant production from
conversion of vacuum energy.  This expansion regime is intrinsically
different from the supercooled inflation regime, since warm
inflation smoothly terminates into a
subsequent radiation dominated regime, without a reheating period.

The warm inflation picture has one immediate conceptual
advantage in that the dynamics is
completely free of questions about quantum-to-classical
transition. The scalar inflaton field is in a
classical state, thus immediately justifying the application
of  a classical evolution equation.  Also, the fluctuations
of the inflaton, which induce the metric perturbations, are
classical.  Furthermore, warm inflation dynamics offers
interesting solutions to the initial
condition problem of inflation \cite{bg},
as well as possibilities for generating cosmic
magnetic fields \cite{bkw}.
 
However despite the conceptual clarity and despite the suggestive
thermodynamic considerations, deriving this dynamics from first
principles quantum field theory is nontrivial.  The key reasons
primarily are technical.  To clarify this
point,  a comparison with supercooled inflationary
dynamics is useful.  In supercooled inflation,
the process of inflation and radiation production are
neatly divided into two different epochs, whereas in
warm inflation dynamics, both processes occur concurrently.
As such, for warm inflation dynamics there is considerable
and nontrivial interplay between the equations of background 
inflationary expansion and quantum field theory dynamics, making it
technically more difficult to solve than supercooled
inflation.  In effect, warm inflation solutions are of an
``all-or-nothing'' type in that if a solution works, it solves
everything and if something fails, the whole solution
becomes useless.  On the other hand,
supercooled inflation solutions are of a ``pick-and-choose''
type, in that every aspect of the problem is compartmentalized,
i.e. inflation, reheating, quantum-to-classical transition,
and there is little continuity amongst the different problems.

Statements have been made about the impossibility of warm inflation
dynamics \cite{yl}. However the dynamical considerations
leading up to these conclusions were limited in their scope,
as had been noted previous to this work \cite{bgr}.
In particular, these works looked for high temperature
warm inflation solutions, under rigid adiabatic, equilibrium
conditions.  Nevertheless, within this limited framework, 
one type of warm inflation solution
was obtained \cite{bgr2,abnpb}, 
and due to the "all-or-nothing" nature
mentioned above, this solutions
can not be discarded as a serious contender
in any more complete theory of inflation \cite{bk}.
Moreover, these early works \cite{yl,bgr} have explicated
one very important point, that warm inflation dynamics is
not trivial and before it can be directly solved, several
missing gaps in the knowledge of dissipative dynamics
must be clarified.

As one step in this direction to fill the missing gaps, recently we
studied the zero temperature dissipative
dynamics of interacting scalar field 
systems in Minkowski spacetime \cite{br}.
This is useful to understand, since
the zero temperature limit constitutes a baseline effect that
will be prevalent in any general statistical state.
What our results show is that
for a broad range of cases, involving interaction
with as few as one or two fields, dissipative regimes are found
for the scalar field system.  This is important for inflationary
cosmology, since it suggests that dissipation
may be the norm not exception for an interacting scalar field system,
thus suggesting that warm
inflation could be a natural dynamics once proper treatment of
interactions is done.

Our analysis of dissipative dynamics starts with the general
Lagrangian,
\begin{eqnarray} 
{\cal L} [ \Phi, \chi_j, \bar{\psi}_k, \psi_k] &=&  
\frac{1}{2} 
(\partial_\mu \Phi)^2 - \frac{m_\phi^2}{2}\Phi^2 - 
\frac{\lambda}{4 !} \Phi^4  
+ \sum_{j=1}^{N_{\chi}} \left\{ 
\frac{1}{2} (\partial_\mu \chi_{j})^2 - \frac{m_{\chi_j}^2}{2}\chi_j^2 
- \frac{f_{j}}{4!} \chi_{j}^4 - \frac{g_{j}^2}{2} 
\Phi^2 \chi_{j}^2  
\right\} 
\nonumber \\ 
&+& \sum_{k=1}^{N_{\psi}}   
\bar{\psi}_{k} \left[i \not\!\partial - m_{\psi_k} -h_{k,\phi} \Phi 
- \sum_{j=1}^{N_\chi} h_{kj,\chi} \chi_j \right] \psi_{k} 
\: , 
\label{Nfields} 
\end{eqnarray} 
with $\Phi \equiv \varphi + \phi$ such that
$\langle \Phi \rangle = \varphi$.  Our aim is to
obtain the effective equation of motion for $\varphi(t)$ and
from that determine the energy dissipated from the $\varphi(t)$
system into radiation.

Using the tadpole method \cite{tadpole}, which requires 
$\langle \phi \rangle =0$,
the effective equation of motion for $\varphi(t)$ emerges
\begin{eqnarray}
\ddot{\varphi}(t) + m_\phi^2 \varphi(t) + \frac{\lambda}{6} \varphi^3(t) 
+\frac{\lambda}{2} \varphi(t) \langle \phi^2 \rangle 
+\frac{\lambda}{6} \langle \phi^3 \rangle
+\sum_{k=1}^{N_{\psi}} h_{k} \langle \bar{\psi_k} \psi_k \rangle= 0 \;.
\label{eeom}
\end{eqnarray}
The field expectation values in this equation are
obtained by solving the coupled set of field equations.
In our calculation, we have evaluated them in a perturbative
expansion using dressed Green's functions \cite{br,lawrie1,bgr}. 
One general feature of these expectation values
is they will depend of the causal history of $\varphi(t)$,
so that Eq. (\ref{eeom}) is a temporally nonlocal equation
of motion for $\varphi(t)$.  

{}Formally, we can examine Eq. (\ref{eeom})
within a Markovian-adiabatic approximation, in which
the equation of motion is local in time and the motion
of $\varphi(t)$ is slow.  At $T=0$, such an approximation
is not rigorously valid.  Nevertheless, this approximation
allows understanding the magnitude of dissipative effects.
{}Furthermore, we have shown in \cite{br} that the nonlocal
effects tend to filter only increasingly higher
frequency components of $\varphi(t)$
from nonlocal effects increasingly further back in time.  
Thus for low
frequency components of $\varphi(t)$, memory only is
retained to some short interval in the past.   Since within
the adiabatic approximation, $\varphi(t)$ only has low frequency 
components, we believe the Markovian-adiabatic
approximation is legitimate at least for order of
magnitude estimates. Within this approximation,
the effective equation of motion for $\varphi(t)$ has the general
form
\begin{eqnarray} 
&& \ddot{\varphi}(t) + m_{\phi}^2 \; \varphi (t) +  
\frac{\lambda}{6} \varphi^3 (t) + \eta (\varphi)  
\dot{\varphi} (t) =0\;, 
\label{final1} 
\end{eqnarray} 
where explicit expressions for the dissipative coefficient $\eta$ 
for various cases are given in \cite{br}.

Based on this equation, energy production will be estimated 
here with full details given in \cite{br}.  Our primary
interest is in the overdamped regime 
\begin{equation} 
m^2(\phi) = m_{\phi}^2 + \lambda \varphi^2/2 < \eta^2,
\end{equation} 
since this is the regime ultimately of interest to warm inflation.
In this regime,
the energy dissipated by the scalar field goes into  
radiation energy density $\rho_r$ at the rate
\begin{equation} 
{\dot \rho}_r = -\frac{dE_{\phi}}{dt} = 
\eta(\varphi) {\dot \varphi}^2 .
\end{equation} 

In \cite{br} we have determined radiation production for
two cases 
\begin{eqnarray}
{\rm (a). } \hspace{0.1cm} & & m(\varphi) > m_{\chi} > 2m_{\psi} \nonumber\\
{\rm (b). } \hspace{0.1cm} & & m_{\chi} > 2 m_{\psi} > m(\varphi).  
\end{eqnarray}
To focus on
a case typical for inflation, suppose the potential
energy is at the GUT scale $V(\varphi)^{1/4} \sim 10^{15} {\rm GeV}$
and we consider the other parameters in a regime consistent
with the e-fold and density fluctuation
requirements of inflation.  Note, although this is a flat nonexpanding
spacetime analysis, since the dissipative effects will
be at subhorizon scale, one expects these estimates to give a reasonable
idea of what to expect from a 
similar calculation done in expanding spacetime.
Expressing the radiation in terms of a temperature scale as
$T \sim \rho_r^{1/4}$, we find for case (a)
$1 {\rm GeV} < T < 10^7 {\rm GeV} < H$ and for case (b)
$T \stackrel{>}{\sim} 10^{14} {\rm GeV} > H$,
where $H = \sqrt{8 \pi V/(3m_p^2)}$.

It should be clarified that the results found in this paper in no 
way require supersymmetry, although they
easily could be applied in SUSY models.  
{}For such models, the low $T$ warm inflation 
solutions suggested by case (a)
could be useful in avoiding gravitino overproduction \cite{tl}.
On the other hand, case (b) in general seems more
interesting, since it offers a very robust
possibility for radiation production. 
Although this is what the formal calculation indicates,
we believe at this point a deeper understanding of
radiation production is necessary.

In regards the potential implications of 
the results discussed in this talk to 
inflationary cosmology, 
we infer that under generic circumstances the scalar inflaton field 
will dissipate a nonnegligible amount of radiation during 
inflation.  In particular,  
the lower bound suggested by the above estimates  
already are sufficiently 
high to preclude a mandatory requirement 
for a reheating period.
Moreover, the high temperature results of case (b) suggest
that warm inflation could be very robust.
Verification of these 
expectations requires a proper extension of these
calculations to expanding spacetime, and within a nonequilibrium
formulation \cite{lawrie2}, which we plan to examine.

\medskip
\noindent
{\bf Acknowledgments}

This work was supported by a PPARC Advanced Fellowship.

\bigskip

\nopagebreak

\end{document}